\documentclass[conference]{IEEEtran}

\IEEEoverridecommandlockouts
\usepackage{cite}
\usepackage{amsmath,amssymb,amsfonts}
\usepackage{algorithmic}
\usepackage{textcomp}

\usepackage{url}
\usepackage{amsmath}
\usepackage{xspace}
\usepackage{booktabs}
\usepackage{subcaption} 

\usepackage[dvipdfmx]{graphicx}

\usepackage{hyperref}
\usepackage{setspace}

\graphicspath{{images_v/}}

\def\Sibling{\mathcal{S}}

\def\Scheme{\textit{S}}
\def\deg{{\it deg}}

\makeatletter
\def\@IEEEpubidpullup{8\baselineskip}
\makeatother

\usepackage{fancyhdr}
\usepackage{kantlipsum}
 \fancypagestyle{plain}{
 \fancyhf{}
 \fancyhead[C]{Conference on \LaTeX} 
 
 }
\usepackage{eso-pic}

\begin{document}

\AddToShipoutPictureBG*{
\AtPageUpperLeft{
\setlength\unitlength{1in}
\hspace*{\dimexpr0.5\paperwidth\relax}
\makebox(0,-0.75)[c]{\textbf{2023 IEEE/ACM International Conference on Advances in Social
Networks Analysis and Mining (ASONAM)}}}}



\title{User's Position-Dependent Strategies in Consumer-Generated Media with Monetary Rewards}

\author{
\IEEEauthorblockN{Shintaro Ueki}
\IEEEauthorblockA{
\textit{Department of Computer Science and}\\
\textit{Communications Engineering,}\\
\textit{Waseda University.}
Tokyo, Japan\\
s.ueki@isl.cs.waseda.ac.jp}
\and
\IEEEauthorblockN{Fujio Toriumi}
\IEEEauthorblockA{
\textit{Department of System Innovation,}\\
\textit{The University of Tokyo.}\\
Tokyo, Japan\\
tori@sys.t.u-tokyo.ac.jp}
\and
\IEEEauthorblockN{Toshiharu Sugawara}
\IEEEauthorblockA{
\textit{Department of Computer Science and}\\
\textit{Communications Engineering,}\\
\textit{Waseda University.}
Tokyo, Japan\\
sugawara@waseda.ac.jp}
}

\maketitle

\begin{abstract}
Numerous forms of consumer-generated media (CGM), such as social
networking services (SNS), are widely used. Their success relies on
users' voluntary participation, often driven by psychological
rewards like recognition and connection from reactions by other
users. Furthermore, a few CGM platforms offer monetary rewards to
users, serving as incentives for sharing items such as articles,
images, and videos. However, users have varying preferences for
monetary and psychological rewards, and the impact of monetary
rewards on user behaviors and the quality of the content they post
remains unclear. Hence, we propose a model that integrates some
monetary reward schemes into the SNS-norms game, which is an
abstraction of CGM. Subsequently, we investigate the effect of each
monetary reward scheme on individual agents (users), particularly in
terms of their proactivity in posting items and their quality,
depending on agents' positions in a CGM network. Our experimental
results suggest that these factors distinctly affect the number of
postings and their quality. We believe that our findings will help
CGM platformers in designing better monetary reward schemes.
\end{abstract}

\begin{IEEEkeywords}
Social media, Consumer-generated media, Monetary reward, Agent-based simulation, co-evolution.
\end{IEEEkeywords}

\section{Introduction}
{\em Consumer-generated media} (CGM), such as {\em social networking
  services} (SNS) and review sites, have become integral to
society. They serve not only as communication
tools~\cite{moorhead2013new,young2014social} but also play crucial
roles in education~\cite{tess2013role,makki2021use},
marketing~\cite{hutter2013impact,hajli2014study,dehghani2016evaluating},
political communication~\cite{Starke2020,Conway2015}, and election
campaigns~\cite{dalton2012effect}. Contrary to traditional media
outlets like television and newspapers, which disseminate information
unilaterally from a few companies or government entities, CGM
flourishes owing to the voluntary participation of its users. This
fact seems unreasonable because voluntarily posting items, such as
articles, images, and videos, to the media incurs, more or less,
psychological, financial, and temporal costs. Thus, it is natural to
assume that there is some incentive for most users to continue their
participation. Thus, it is crucial to analyze the effect of different
types of incentives to attract more users.
\par

In general, the common incentives for contributing to CGM are
psychological rewards for fulfilling their need for belonging,
self-expression, and self-approval~\cite{Nad2012} in virtual
connections on the Internet. To realize these, numerous CGMs include
``Like'' buttons, comments, stamps, and other features that present
responses among users. Moreover, some media offer a few types of
engagements that provide opportunities for monetary rewards, such as
reward program points, advertising revenues, or sponsorship deals with
well-known brands as incentives for participating to post more quality
items. For example, on YouTube 
(\url{https://www.youtube.com/}), users with a large subscriber base
can be financially rewarded by showing advertisements. Rakuten Recipe
(\url{https://recipe.rakuten.co.jp/ }), a website for sharing cooking
recipes and cooking experiences of those recipes, allows users to earn
points by posting recipe articles that can be used to make purchases
at the Rakuten e-commerce site. However, it remains unclear which
incentives truly influence user behavior.
\par

Numerous studies have attempted to identify the reason people actively
participate in CGMs from different viewpoints, such as survey and
empirical investigations~\cite{hosen2021,KHAN2017}, social network
analysis~\cite{chalakudi2023,Pazdor2022} and (evolutionary) game
theory~\cite{Wan2023,Gao2018,Toriumi2012MetaReward,Usui2022Scheme}. Our
study has adopted an evolutionary game-theoretic approach because we
believe that it is suitable for a formal analysis of how user
behavioral strategies are influenced by the incentives provided by CGM
and their positions within social networks. For example, Toriumi et
al.\cite{Toriumi2012MetaReward} proposed the {\em meta-reward game}
based on {\em public goods game}~\cite{AxelrodMetaNorms} and then
analyzed requirements for a balance between costs and rewards to
ensure that items are constantly posted. Hirahara et
al.\cite{Hirahara2014SnsNorms} extended this game, called {\em
  SNS-norms game}, by adding simple responses such as ``Like'' button
and ``read'' 
marks. They suggested that these low-time cost responses increase user
activities. Moreover, to investigate the effect of monetary rewards on
the number and quality of posted items, Usui et
al.~\cite{Usui2022Scheme} introduced a few schemes for monetary
rewards to the SNS-norms game. They also show that by
appropriately providing monetary rewards, users attempt to improve the
quality of the items they post. Although users' behavioral strategies
in a CGM are partly determined based on their positions in the
networks, such as normal users and influencers with many followers,
these studies overlook the users' positions in the social networks
because they used the conventional {\em genetic algorithm} (GA) on a
CGM network.
\par

Therefore, we investigate how users' behaviors vary depending on their
positions when a monetary reward scheme is introduced in CGM. For this
purpose, we use the {\em multiple-world genetic algorithm}
(MWGA)~\cite{Miura2021MwgaNorms}, which is a {\em co-evolutionary}
algorithm by extending the conventional GA, to ensure that all agents
(users) can examine different behavioral strategies with neighbors
that also have various strategies in the same network structure. We
also use the {\em connecting nearest neighbor} (CoNN)
network~\cite{CNN} as a CGM to investigate the difference in agents'
behaviors, because the CoNN network is generated based on triadic
closure, that is, triangles in the network tend to close as links form
between friends of friends~\cite{easley_kleinberg_2010} and have the
common properties of complex networks, that is, the properties of
small-world, scale-freeness, and high-cluster coefficients. We also
assume that posting high-quality an item incurs a higher cost owing to
the need for a more engaging topic and investing in detailed
elaboration, but these posts can attract more comments. Subsequently,
we experimentally demonstrate the manner in which agent strategies
concerning the posting/comment rates and the quality of their items
vary depending on their degrees, that is, the number of
friends/followers, and on agents' preferences for monetary or
psychological reward. We believe that our findings on the effects on
agents' behaviors will help in the design of future CGMs to improve
item quality by adopting a monetary rewards scheme.
\par

\section{Related Work}
\label{related}
Studies on CGMs have been widely conducted. For instance, to
empirically analyze specific users' behaviors and reactions, Berry et
al.~\cite{textmine} analyzed structures of hashtags using text mining
techniques, and Marett et al.~\cite{snowball} investigated users'
reactions using a snowball sampling method. Dhir et
al.~\cite{Dhir2019} identified that psychological self-efficacy and
the users' habits affect the use of the Facebook ``Like'' button from
the perspective of {\em behavioral intention theory}. Several studies
focused on the influence of monetary rewards on users' behaviors in
CGM. Cvijikj et al.~\cite{Cvijikj2013} found that
monetary rewards for users on Facebook actively increased the number
of comments, while it led to a decrease in passive responses such as
the use of "Like" buttons. Chen et al.~\cite{Chen2019}
analyzed online inventor communities. They discovered that providing
monetary incentives would increase the diversity of content within the
media and result in more interest from the community members but does
not lead to more stock recommendations.
\par

Apart from these empirical studies, there are studies from an
evolutionary game theoretic perspective to identify the behaviors of
users in
CGMs~\cite{Gao2018,Wan2023,Toriumi2012MetaReward,Hirahara2014SnsNorms,Usui2022Scheme}. For
example, Gao et al.~\cite{Gao2018} introduced an information
disclosure game and demonstrated that social media could promote the
regional government to disclose information during environmental
incidents through the top-down intervention and bottom-up reputation
mechanisms. Wan et al.~\cite{Wan2023} proposed an information
diffusion model using evolutionary game theory, and analyzed two
aspects that may affect the diffusion by identifying the phenomena
appearing in information dissemination between individuals and the
general environment on online social networks. Toriumi et
al.~\cite{Toriumi2012MetaReward} found that CGMs share characteristics
with public goods game~\cite{AxelrodMetaNorms} and proposed the {\em
  meta-reward game}, which is a dual problem with Axelrod's {\em
  meta-norms game}. Subsequently, they showed that meta-comment, that
is, comment-return or comment on the comment, plays a crucial role in
incentivizing voluntary activities. Subsequently, the meta-reward game
was modified to {\em SNS-norms game}~\cite{Hirahara2014SnsNorms} to
fit the nature of CGM and further extended to {\em SNS-norms game with
  monetary reward and article quality} (SNS-NG/MQ) to investigate the
effect of monetary rewards by identifying the common dominant
behavioral strategy for all agents. However, unlike ours, they
overlooked the diverse strategies depending on the positions of users
in the network.

\section{Preliminary}
\subsection{SNS-norms game and agent network}
The SNS-norms game~\cite{Hirahara2014SnsNorms} is an abstract model
for the agents' behaviors with strategies in an SNS/CGM. Let graph
$G=(A,E)$ be a CGM network, where $A=\{a_1, a_2, \dotsc ,a_N \}$ is
the set of $N$ agent nodes (or simply agents), each of which
corresponds to a user in the CGM, and $E$ is the set of edges
connecting two agents. Thus, edge $(a_i, a_j)\in E$ represents the
connection, referring to the relationship between agents such as
friends. We denote the neighboring agents of $a_i$, $N_{a_i}=\{a_j\in
A \mid (a_i,a_j)\in E\}$. Each agent $a_i$ has two parameters, posting
rate $B_i$, comment/meta-comment rate $L_i$, ($0\leq B_i, L_i \leq
1$), to control its behavioral strategy. These parameter values
evolved through the interaction with neighboring agents in $G$ using
GA such that they can gain more fitness values, which are the total
rewards they earn.
\par

\begin{table}
 \caption{Parameters in SNS-norms game}
 \label{tab:SNS-norms}
 \centering \small
 \begin{tabular} {ll}
   \toprule Meaning & Parameter \\ \midrule Cost of an item post &
   $F\geq 0$ \\ Reward by viewing an item & $M\geq 0$ \\ Cost of a
   comment & $C\geq 0$ \\ Reward by viewing a comment & $R\geq 0$
   \\ Cost of a comment return & $C''\geq 0$ \\ Reward by viewing a
   comment return & $R''\geq 0$ \\ \bottomrule
 \end{tabular}
\end{table}

We describe the outline of the SNS-norms game. The parameter used here
is listed in Table~\ref{tab:SNS-norms}. At the turn of agent $a_i\in
A$, the value of parameter $S$ ($0\leq S\leq 1$) is randomly selected;
$S$ represents, for example, the level of interest of the content of
the next item, and the value of $B_i$ indicates the extent to which
$a_i$ actively post the item. Thus, agent $a_i$ determines whether it
posts an item or not with parameter $B_i$. If $S < 1-B_i$, $a_i$ does
not post an item, and the game round ends; otherwise, $a_i$ posts an
item with cost $F$. Another agent $a_j\in N_{a_i}$ obtains reward $M$
by viewing the posted item. Subsequently, $a_j$ comments on the item
with probability $L_j$ and incurs cost $C$; if it does not, the
behavior of $a_j$ ends with only viewing and no comment. Subsequently,
$a_i$ who posted the first item gains reward $R$ through the comment
from $a_j$. When $a_i$ responds to it, that is, posting a
meta-comment, with probability $L_i$. If $a_i$ posts it, $a_i$ incurs
cost $C''$ and $a_j$ obtains reward $R''$. Note that all rewards
described here are psychological rewards and no physical materials or
monetary rewards are provided.
\par

\subsection{Multiple-world GA}\label{mwga}
Previous studies have used naive GA to determine the dominant strategy
common for all agents that yield high rewards in a CGM. However,
agents' optimal strategies also depend on their positions in the
network and the strategies of neighboring agents that also evolve
concurrently to fit their neighbors, including the first agents as
well as the agents neighboring on the other sides. We employ MWGA to
determine agents' strategies; we will describe it in brief.
\par

MWGA is an extension of the GA to facilitate co-evolutionary
learning within a network of agents. It allows agents in
MWGA to explore various strategies in parallel worlds that are
replicated.
First, in MWGA, $W$ ($\geq 1$) networks, $G^l=(A^l,E^l)$
(for $1\leq l\leq W$), are duplicated from $G$; therefore, $A^l=A$ and
$E^l=E$. For agent $a_i$ in $G$, the set of its copies (clones) is
denoted by $\Sibling_i=\{a_i^1, \dots, a_i^W\}$ whose element is
called a {\em sibling agent}. Assuming that sibling agent $a_i^l$ has
its strategy in the $l$-th world $G^l$, all sibling agents of $a_i$
experience interactions with their neighboring sibling agents, which
also have different strategies at the same relative position. This
suggests that each $a_i^l$ has different values of posting rate
$B_i^l$ and comment/meta-comment rate $L_i^l$, which are expressed by
3-bit binary genes, ranging from $0/7$ to $7/7$.
\par

Similar to the conventional GA, MWGA consists of three stages;
(parents) selection, crossover, and mutation, to create genes for the
next generation, but its selection stage is distinctive. Unlike GA, in
which the gene of $a_i$ is likely to be inherited from its and
neighboring agents, the gene of $a_i^l$ ($1\leq l\leq W-1$) in MWGA is
inherited from $\Sibling_i$, which had another experience at the
same position, according to the values of the fitness function. 
MWGA enables agents to learn location-specific strategies 
because each agent evolves using its experience as $\Sibling_i$.
Moreover, the $W$-th world, $G^W$, is a test world to
confirm the optimal strategies of all agents that are usually
selected from the different worlds. The detailed encoding in our model
will be explained in Section~\ref{game}.
\par

\section{Proposed Model and Methodology}\label{model}
We propose the {\em SNS-norms game with monetary reward and article
  quality} (SNS-NG/MQ) to investigate the effect of monetary rewards
on the behaviors of agents and the quality of the item they post,
which are article, image, or video content. The main differences from
the previous model~\cite{Usui2022Scheme,Usui2021Monetary} are that it
did not consider the uniqueness of its standpoints and differences in
behavioral strategies of neighbors. Therefore, in the previous model,
genes were inherited from the agents with different standpoints. This
seems acceptable when all agents are uniform like a complete graph,
but the actual network structures are far from the complete
graph. Thus, we extended their model to ensure that it can be used
with all type of networks. Further, we eliminated our unnecessary
agent types in our model because our main goal is to examine
differences in behavioral strategies with standpoints, that is, places
in the network, 
and unnecessary agent types hinder this analysis. Specifically, the
main reason for introducing such unnecessary agents in previous
studies was that they assumed a CGM in which a large proportion of
agents do not post an item and only view it. 
Our research does not need to take this into account supecifically, 
since agents in positions where such behaviour would be optimal can also converge on their unique strategies.
Our contribution enables
us to determine the most effective monetary reward scheme for agents
at the places. Conversely, as CGM managers also aim to encourage
activity from specific types of agents, they can design reward schemes
that are appropriate for these target groups. 
\par

\begin{figure}[bt]
\centering
\includegraphics[width=\linewidth]{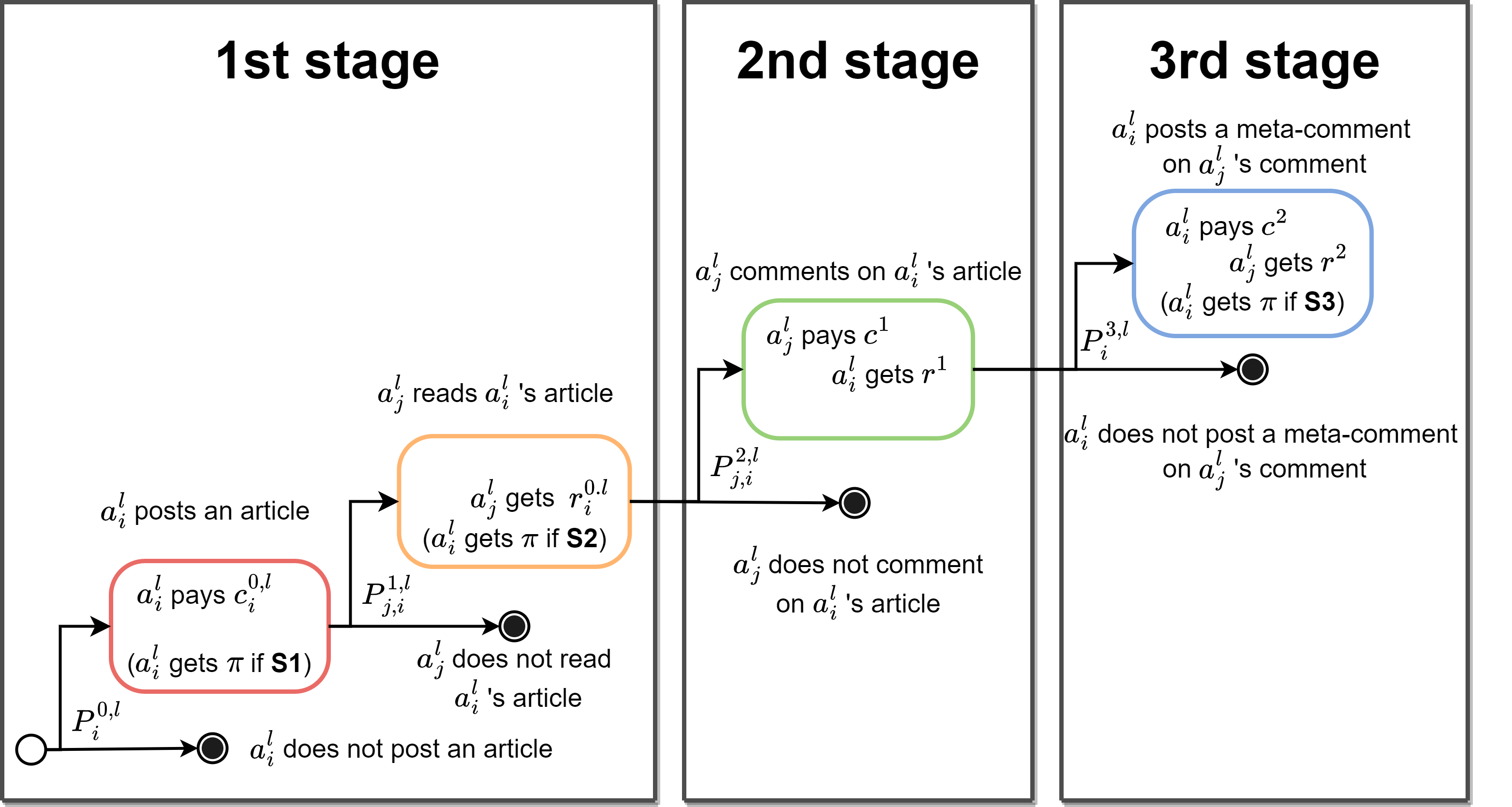}
\caption{One game round of SNS-NG/MQ}
\label{fig:game}
\end{figure}

\subsection{Agents in network}
As in the SNS-norms games, agents have interactions with neighboring
agents in $G=(A,E)$. In addition to the behavioral parameters $B_i$
and $L_i$ for $a_i\in A$, we introduce another parameter $Q_i$, ($0<
Q_{\it min}\leq Q_i\leq 1$), denoting the item
quality~\cite{Usui2022Scheme}, relating to the willingness of agent
$a_i$ to spend to improve the quality of their items. We presume that
quality items are more likely to receive comments and this reduces
posting frequency because of the effort they involve. Note that
$Q_{\it min}$ is the minimum value of quality and, unlike $B_i$ and
$L_i$, $Q_i>0$.
\par

We also introduce parameter $M_i$ ($0\leq M_i\leq 1$) to express the
preference for monetary reward. As we believe that this preference is
intrinsic and constant, we set it as a constant value after it is
defined initially for $a_i$. Subsequently, agents are classified into
the following two types of agents:
\begin{equation}
V_\alpha =\{a_i \in A \mid M_i < 0.5 \}, V_\beta =\{a_i \in A \mid M_i \geq 0.5 \}
\end{equation}
Thus, $V_\alpha$ and $V_\beta$ are the sets of agents that relatively
prefer either psychological or monetary rewards, respectively.
\par

As we consider cases where agents are in various places in a CGM
network, let graph $G$ be a CoNN network, which is an undirected
complex network based on the {\em connecting nearest neighbor}
model~\cite{CNN}. A CoNN network is built based on the basic property,
friends of friends tend to be direct friends owing to the homophily of
friends. Thus, it is generated by probabilistically conducting the
following two steps, {\em agent addition} and {\em edge addition},
until $|A|=N>0$. Initially, $G=(A,E)$ is a complete graph of one or a
few agents.
\par

\subsubsection{Agent addition} A new agent $a$ is added to $A$ with
probability $1-u$ and is connected to the agent $a'$ selected from
$A\setminus\{a\}$ randomly with an actual edge. Moreover, $a$ is
connected to each agent in $N_{a'}$ with a potential edge.

\subsubsection{Edge addition} A potential edge is randomly selected
and converted to an actual edge with probability $u$.
\par

Note that $u$ ($0 \leq u \leq 1$) is called the {\em conversion probability}.

\subsection{SNS-norms game with monetary reward and article quality}\label{game}
Agent $a_i^l$ on $G^l=(A^l,E^l)$ ($1\leq l \leq W$) performs the
SNS-NG/MQ with its neighboring agents. It is similar to the process
involved in the SNS-norms game but is modified to consider the quality
of items and monetary rewards. One game round is illustrated in
Fig.~\ref{fig:game}, in which three types of monetary reward schemes,
\Scheme1, \Scheme2, and \Scheme3, are shown, but only one of them is
implemented for each game episode~\cite{Usui2021Monetary}. In these
schemes, monetary reward $\pi$ ($\geq 0$) is offered to $a_i^l$ when:
\begin{itemize}
\item[\Scheme1:] Agent $a_i^l$ has posted an item
\item[\Scheme2:] The item posted by $a_i^l$ is viewed
 by a neighboring agent; and
\item[\Scheme3:] Agent $a_i^l$ returns a meta-comment to the comment from one of its neighboring agents.
\end{itemize}
Therefore, it is feasible for $a_i^l$ to earn multiple $\pi$ in
\Scheme2 and \Scheme3. We assume that monetary rewards are provided by
the platformer of the CGM. Note that $\pi=0$ means that no monetary
rewards are adopted; this scheme is denoted by \Scheme0.
\par

We denote the parameters to determine the behaviors of $a_i^l$ by
$B_i^l$, $L_i^l$, and $Q_i^l$. A game round of $a_i^l$ with the
neighboring agent in $N^l_{a_i^l}\subset A^l$ proceeds in the three
stages as follows. In the first stage, agent $\forall a_i^l\in A^l$
posted an item with probability $P^{0,l}_i$, where
\begin{equation}
\label{eq:post}
  P_i^{0,l}=B_i^l \times\dfrac{Q_{\it min}}{Q_i^l}.
\end{equation}
Therefore, $P_i^{0,l}$ indicates that agents have fewer opportunities
for posting quality items. The game round is conducted concurrently,
that is, all agents perform the first stage concurrently. When $a_i^l$
posts the item, it pays an amount $c_i^{0,l}$. The definition of
various types of costs will be described later. If $a_i^l$ does not
post an item this round of $a_i^l$ ends. Only when monetary reward
scheme \Scheme1 is adopted, $a_i^l$ receive monetary reward $\pi$
regardless of the quality of the item. Thereafter, agent $\forall
a^l_j\in N_{a_i^l}^l$ may discover and view $a^l_i$'s posted item with
probability $P^{1,l}_{j,i} = Q_i^l/s_j^l$, where $s_j^l$ is the number
of items posted by agents in $N_{a_j^l}^l$ at the same round;
therefore, $P^{1,l}_{j,i}$ means that items are more likely to be
found when their quality is higher and when the number of posted items
is low. Note that to avoid dividing by zero we define $P^{1,l}_{j,i} =
0$ when $s^l_j=0$. The agent $a^l_j$ who viewed the item gains
psychological reward $r^{0,l}_i$ as a benefit of information
sharing. The definition of rewards is also explained later. If the
adopted monetary reward scheme is \Scheme2, each time the item is
discovered and viewed, monetary reward $\pi$ is provided to $a^l_i$.
\par

In the second stage, agents $a^l_j$ who viewed the item from $a_i^l$
comments on the item with probability $P^{2,l}_{j,i} = L^l_j \times
Q^l_i$; thus, quality items are likely to receive comments. Agent
$a_j^l$ pays cost $c^1$ for comment, and $a^l_i$ obtains psychological
reward $r^1$ for each comment received. In the third stage, $a^l_i$
who posted the item and received the comment from $a^l_j$ can respond
to a meta-comment from $a_j^l$ with probability $P^{3,l}_{i} = L^l_i
\times Q^l_i$. Subsequently, $a^l_i$ pays an amount of $c^2$ for the
meta-comment and $a^l_i$ gains a psychological reward
$r^2$. $P^{2,l}_{j,i}$ and $P^{3,l}_{i}$ also indicate that
higher-quality items posted by $a^l_i$ are likely to receive a
meta-comment. Furthermore, if scheme \Scheme3 is adopted, $a_i^l$
receives monetary reward $\pi$ when $a_i^l$ posts a meta-comment. The
game flow mentioned above is called one {\em game round} of $a_i^l\in
A^l$ and a round for all agents is called a {\em game}.
\par

\subsection{Cost, reward, and utility in game}
By referring to the study by Okada et al.~\cite{okada2015}, the values of the costs and psychological rewards are defined as follows:
\begin{equation}
\label{eq:cost}
\begin{array}{ll}
c^{0,l}_i=c_{\it ref}\times Q^l_i 	&r^{0,l}_i=c^{0,l}_i\times \mu\\ 
c^1 =c_{\it ref}\times \delta		&r^1=c^1\times \mu\\
c^2 =c^1\times \delta		&r^2=c^2\times \mu\\
\end{array}
\end{equation}
where $c_{\it ref} > 0$ is the reference value for cost and
psychological reward, $\mu$ is the ratio of cost to psychological
reward, and $\delta$ ($0\leq \delta \leq 1$) is the reduction ratio of
cost/reward between game stages. Therefore, if $Q_i$ is high, the cost
for posting an item $c^{0,l}_i$ and the reward for viewing an item
$r^{0,l}_i$ also becomes high. In contrast, the cost for comment and
meta-comment $c^1$, $c^2$ and psychological reward $r^1$, $r^2$ do not
have a relationship to $Q_i$.
\par

Utility $u^l_i$ of $a_i^l$ in a single game are then defined by
\begin{equation}
\label{eq:utility}
u_i^l=(1-M_i)\times R^l_i + M_i\times K^l_i - C^l_i,
\end{equation}
where $R^l_i, K^l_i,$ and $C^l_i $ are the summed psychological
rewards, monetary rewards, and cost provided to/incurred $a_i^l$ in
the game. This utility is used as the fitness value in MWGA, and thus,
all agents attempt to increase their utilities.
\par

\subsection{Multiple-world GA for SNS-NG/MQ}\label{mwga-MQ}
The co-evolutionary learning algorithm, MWGA, is applied such that
each agent evolves its behavioral strategy specified by $B_i$, $L_i$
and $Q_i$ (more precisely, $B_i^l$, $L_i^l$ and $Q_i^l$ in the $l$-the
world). Its use for SNS-NG/MQ is almost identical to that for the
SNS-norms game described in Section~\ref{mwga} except $Q_i^l$ that
does not appear in the SNS-norms game. $Q_i^l$ is also encoded by
3-bit gene whose values correspond to $1/8$ ($=Q_{\it min}$), $2/8,
\dots, 8/8$.
\par

One generation is defined as $N_{\it gen}$ ($\geq 1$) games and one
episode consists of $g$ ($\geq 1$) generations. The fitness value
$U_i^l$ of $a_i^l$ used for the selection after each generation is the
sum of $u_i^l$ obtained in the SNS-NG/MQ. The selection, crossover,
and mutation processes are performed as follows:
\par

\subsubsection{Selection}\label{select}
Two parents of each agent $a_i^l\in A^l$ ($1\leq l\leq W-1$) are
itself and one sibling agent in
$\Sibling_i^{-l}=\Sibling_i\setminus\{a_i^l\}$ selected using the {\em
  roulette wheel selection}, that is, agent $a_i^k\in\Sibling_i^{-1}$
is selected with the probability $\Pi_i^{k}$:
\begin{equation}
  \Pi^k_i=\dfrac{(U^k_i-U_{i,{\it
        min}})^2+\varepsilon/(W-1)}{\sum_{v^k_i\in A^{-l}_i}
    (U^k_i-U_{i,{\it min}})^2+\varepsilon},
\end{equation}
where $U_{i,{\it min}}=\min_{v^k_i\in A_i^{-l}}U^k_i$. Value
$\varepsilon$ ($\ll 1$) is a positive value to avoid zero division (we
set $\varepsilon = 0.00001$).
\par

Meanwhile, agent $a_i^W\in A^W$ in the next generation is set to one
of the sibling agents $\Sibling_i$ with the highest utility $U_i$, and
the crossover is not applied. Therefore, the $W$-th world consists of
agents with the highest utilities in all worlds.

\subsubsection{Crossover}\label{cross}
In $l$-th ($1\leq l\leq W-1$) worlds, the gene of the child is
composed of the selected parents using the {\em uniform crossover} of
the selected parents.

\subsubsection{Mutation}
\label{mute}
To prevent falling into a local optimum, it is done by flipping each
bit of the 9-bit gene of each agent in $A^l$ for $1\leq \forall l \leq
W$ with a {\em mutation probability} $m$, where $0 \leq m \leq
1$.
\par

Finally, the outcome in each generation from MWGA is defined as the
results of the $W$-th world.

\begin{table*}
\caption{Parameter values in experiments}
\label{tab:param}
\centering
\begin{tabular}{lll}
 \toprule
 Description & Parameter & Value \\
 \midrule
 Number of agents & $N=|A|$ & 400 \\
 Number of agents preferring psychological reward & $|V_\alpha|$ & 200 \\
 Number of agents preferring monetary reward & $|V_\beta|$ & 200 \\
 Reference value for cost and psychological reward & $c_{\it ref}$ & 1.0 \\
 Ratio of cost to psychological reward & $\mu$ & 8.0 \\
 Cost ratio between game stages & $\delta$ & 0.5 \\
 Number of worlds in MWGA & $W$ & 10 \\
 Number of games in a generation & $N_{\it gen}$ & 4 \\
 Number of generations (episode length) & $g$ & 1000 \\
 Probability of mutation & $m$ & 0.01 \\
 \bottomrule
\end{tabular}
\end{table*}

\section{Experiments and Discussion}
\subsection{Experimental setting}
We conducted the experiments of SNS-NG/MQ to investigate the agents'
individual and diverse behavioral strategies using MWGA and compared
these results with those by the naive GA as used in the previous
study~\cite{Usui2022Scheme} to investigate the difference when
overlooking the places in a CGM network. We also investigated the
impact of monetary reward schemes, the reward value $\pi$ on agents'
strategies, and the number of posts depending on their places in
the network. In particular, we focused on the degrees of the network
because it shows the number of friends. We used the network based on 
the CoNN model whose conversion rate $u=0.9$ (because this network has
features similar to certain instances of actual
networks~\cite{Hirahara2014SnsNorms}). If nothing is stated,
$\pi=1.0$. The other parameter values are listed in
Table~\ref{tab:param}. These values are taken from the previous
study~\cite{Usui2022Scheme}. Because the result of MWGA is the outcome
of the $W$-th world, we simply denote their parameters such as $B_i$
$(=B_i^W)$, $L_i$ $(=L_i^W)$, and $Q_i$ $(=Q_i^W)$. The averages of
these parameters are denoted by $B$, $L$, and $Q$. The data shown
below are the results of 100 experimental runs.
\par

\begin{figure}[bt]
\centering
\begin{minipage}[t]{0.49\linewidth}
  \centering
  \includegraphics[width=\linewidth]{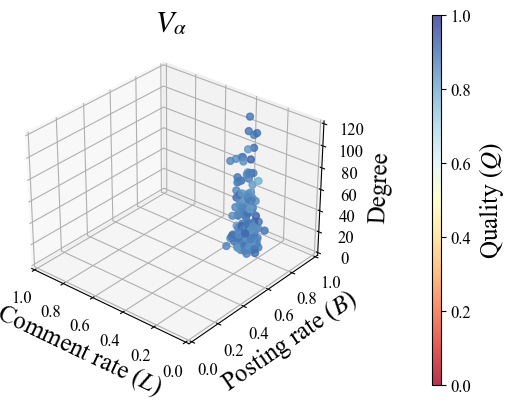}
  \subcaption{\Scheme0 $V_\alpha$}
  \label{fig:S0aGA}
\end{minipage}
\hfil
\begin{minipage}[t]{0.49\linewidth}
  \centering
  \includegraphics[width=\linewidth]{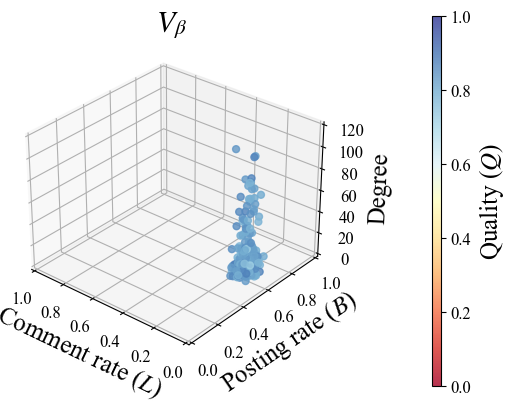}
  \subcaption{\Scheme0 $V_\beta$}
  \label{fig:S0bGA}
\end{minipage}
\caption{Strategy parameters and agent's degree (GA)}
\label{fig:3DGA}
\end{figure}

\begin{figure*}[bt]
\centering
\begin{minipage}[t]{0.24\linewidth}
  \centering
  \includegraphics[width=1.05\linewidth]{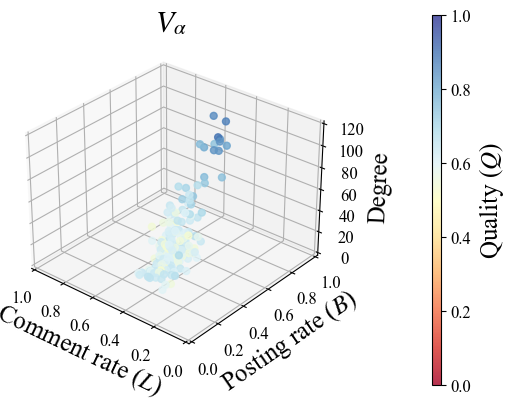}
  \subcaption{\Scheme0 $V_\alpha$}
  \label{fig:S0a}
\end{minipage}
\hfil
\begin{minipage}[t]{0.24\linewidth}
  \centering
  \includegraphics[width=1.05\linewidth]{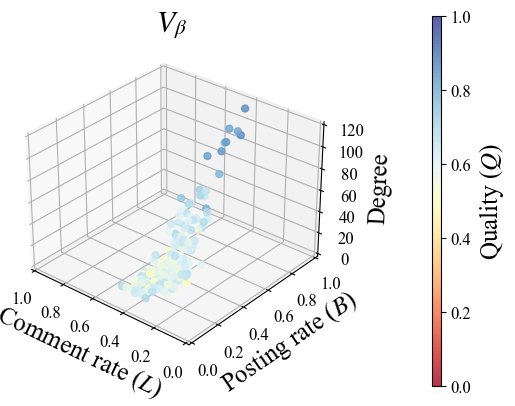}
  \subcaption{\Scheme0 $V_\beta$}
  \label{fig:S0b}
\end{minipage}
\hfil
\begin{minipage}[t]{0.24\linewidth}
  \centering
  \includegraphics[width=1.05\linewidth]{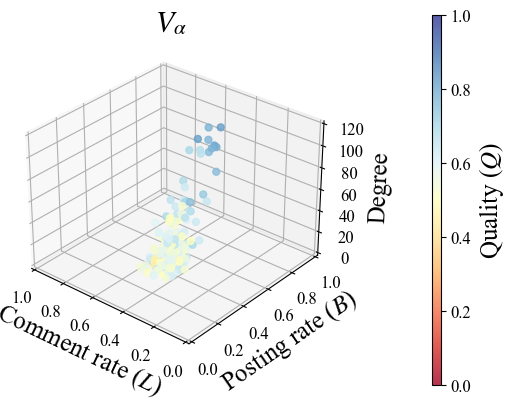}
  \subcaption{S1 $V_\alpha$}
  \label{fig:S1a}
\end{minipage}
\hfil
\begin{minipage}[t]{0.24\linewidth}
  \centering
  \includegraphics[width=1.05\linewidth]{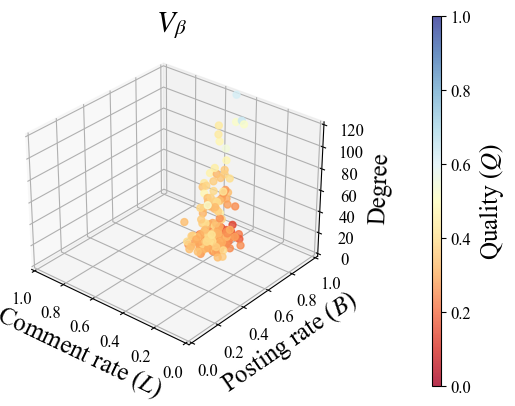}
  \subcaption{S1 $V_\beta$}
  \label{fig:S1b}
\end{minipage}
\hfil
\\
\begin{minipage}[t]{0.24\linewidth}
  \centering
  \includegraphics[width=1.05\linewidth]{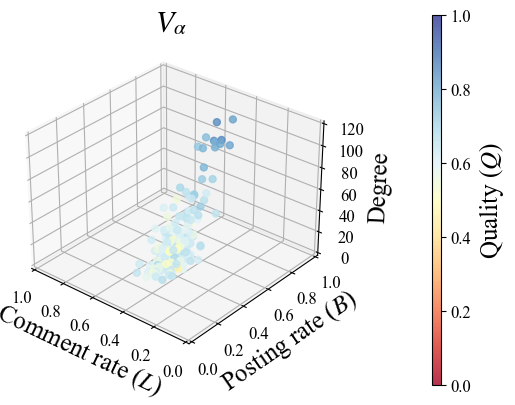}
  \subcaption{S2 $V_\alpha$}
  \label{fig:S2a}
\end{minipage}
\hfil
\begin{minipage}[t]{0.24\linewidth}
  \centering
  \includegraphics[width=1.05\linewidth]{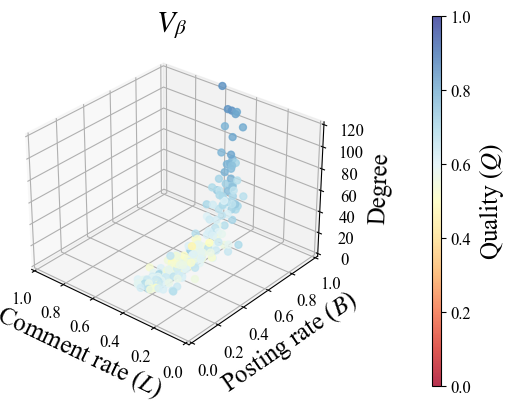}
  \subcaption{S2 $V_\beta$}
  \label{fig:S2b}
\end{minipage}
\hfil
\begin{minipage}[t]{0.24\linewidth}
  \centering
  \includegraphics[width=1.05\linewidth]{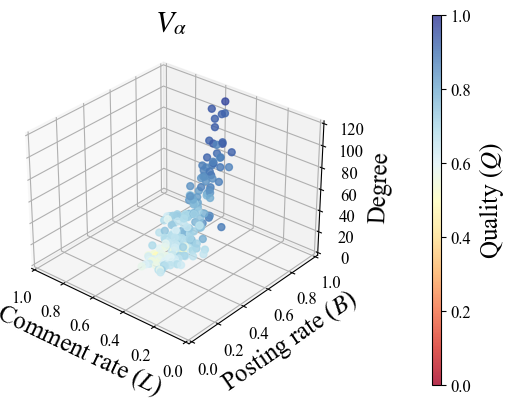}
  \subcaption{S3 $V_\alpha$}
  \label{fig:S3a}
\end{minipage}
\hfil
\begin{minipage}[t]{0.24\linewidth}
  \centering
  \includegraphics[width=1.05\linewidth]{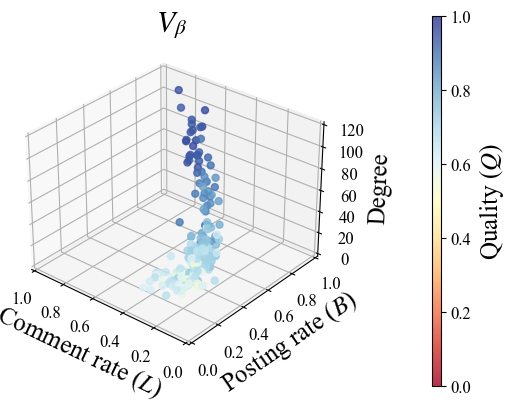}
  \subcaption{S3 $V_\beta$}
  \label{fig:S3b}
\end{minipage}
\hfil
\caption{Distribution of strategy parameters and agent's degree (MWGA)}
\label{fig:3DMWGA}
\end{figure*}

\subsection{Distribution of behavioral strategies with GA and MWGA}
The distribution of behavioral strategies of agents in $V_\alpha$
(preferring psychological reward) and $V_\beta$ (preferring monetary
reward) in scheme \Scheme0 (no monetary reward is provided), when
agents learned their strategies with GA~\cite{Usui2022Scheme}, is
plotted in Fig.~\ref{fig:3DGA}. This figure indicates that agents in
$V_\alpha$ and $V_\beta$ respectively, learned, regardless of their
degrees, the almost identical strategies that were specified by the
values of $B$, $L$, and $Q$. It also shows that agents in $V_\alpha$
were more cooperative (higher $B$, $L$ and $Q$ as shown in
Fig.~\ref{fig:S0aGA}) than those in $V_\beta$ (Fig.~\ref{fig:S0bGA})
because only psychological rewards are the incentive to participate in
a CGM.
\par

The distribution of behavioral strategies of agents learned with MWGA
in scheme \Scheme0, \Scheme1, \Scheme2, or \Scheme3 is plotted in
Fig.~\ref{fig:3DMWGA}. This clearly shows a different trend from that
in Fig.~\ref{fig:3DGA}, that is, their strategies varied according to
their degrees (the number of friends/followers) in the network even
under scheme \Scheme0 and even if agents are of the same
type. Figures~\ref{fig:S0a} and \ref{fig:S0b} show that agents in
$V_\alpha$ and $V_\beta$ have a similar comment rate $L$, and the
agents with high degrees in $V_\alpha$ have a higher posting rate
$B_i$ and higher quality $Q_i$. Agents with lower degrees in $V_\beta$
posted items with a lower rate ($B_i$) and their quality value $Q_i$
was relatively lower. This is because agents with high degrees had
more chances to receive comments from their neighbors, raising the
posting rate $B_i$. They also improved the quality to receive more
comments, which brought more psychological rewards. In contrast,
agents with lower degrees post relatively poor-quality items as few
comments were expected. We believe that these results are intuitively
acceptable, as agents' behaviors are naturally influenced by the
behaviors and numbers of their neighbors, particularly the cooperative
behavior of commenting. Note that we conducted the same experiments
using GA in \Scheme1, \Scheme2, and \Scheme3, but their results were
also uniform; the detailed results were identical to those in the
previous study~\cite{Usui2022Scheme}.
\par

Let us assess the results of behavioral strategies with MWGA when
monetary rewards were adopted. For example, if we compare
Figs.~\ref{fig:S1a} and \ref{fig:S1b} (\Scheme1), agents in $V_\beta$
in \Scheme1 were likely to post more items, but their quality values
$Q_i$ were significantly lower than those of agents in
\Scheme0. Meanwhile, distributions of $B_i$ of agents in $V_\alpha$ in
\Scheme0 and \Scheme1 were almost identical but the values of $Q_i$ of
agents in $V_\alpha$ were slightly lower than those in \Scheme0. In
particular, agents with lower degrees posted more low-quality
items. Therefore, we could say that \Scheme1 provided an incentive for
agents to increase the number of items by decreasing the quality,
particularly those with low degrees preferring monetary rewards
($V_\beta$). Their strategies were reasonable because they could gain
monetary reward $\pi$ only by posting any items. Although the results
are not shown owing to the page limitation, the average psychological
reward and utility of agents in $V_\alpha$ considerably decreased.
\par

Next, we compare Figs.~\ref{fig:S2a} and \ref{fig:S2b} (\Scheme2) with
Figs.~\ref{fig:S0a} and \ref{fig:S0b} (\Scheme0). There is no
significant difference in these schemes except that agents in
$V_\beta$ had slightly higher $B_i$. This is because the poster's
monetary rewards when being viewed are not accompanied by
psychological rewards, only viewer agents gain psychological rewards,
such that only agents in $V_\beta$ are more likely to encourage
posting items. Meanwhile, all agents in $V_\alpha$ and $V_\beta$
increased the values of $B_i$ and $Q_i$ in scheme \Scheme3 as shown in
Figs.~\ref{fig:S3a} and \ref{fig:S3b}. This situation is desirable
because all agents attempt to post more items with better quality. The
monetary reward for meta-comments is tied to the psychological reward
that the first item poster gained from the received comments. If the
quality of an item is high, both monetary and psychological rewards
can be obtained with high probability. This can be an incentive for
posting quality items. Moreover, agents with a high degree in
$V_\beta$ considerably increased $L_i$ over those in $V_\alpha$,
because the responses by meta-comments are only the way to gain
monetary rewards; this is an incentive to increase $L_i$.
\par

\begin{figure}[bt]
\centering
\begin{minipage}[t]{0.95\linewidth}
  \centering
  \includegraphics[width=0.95\linewidth]{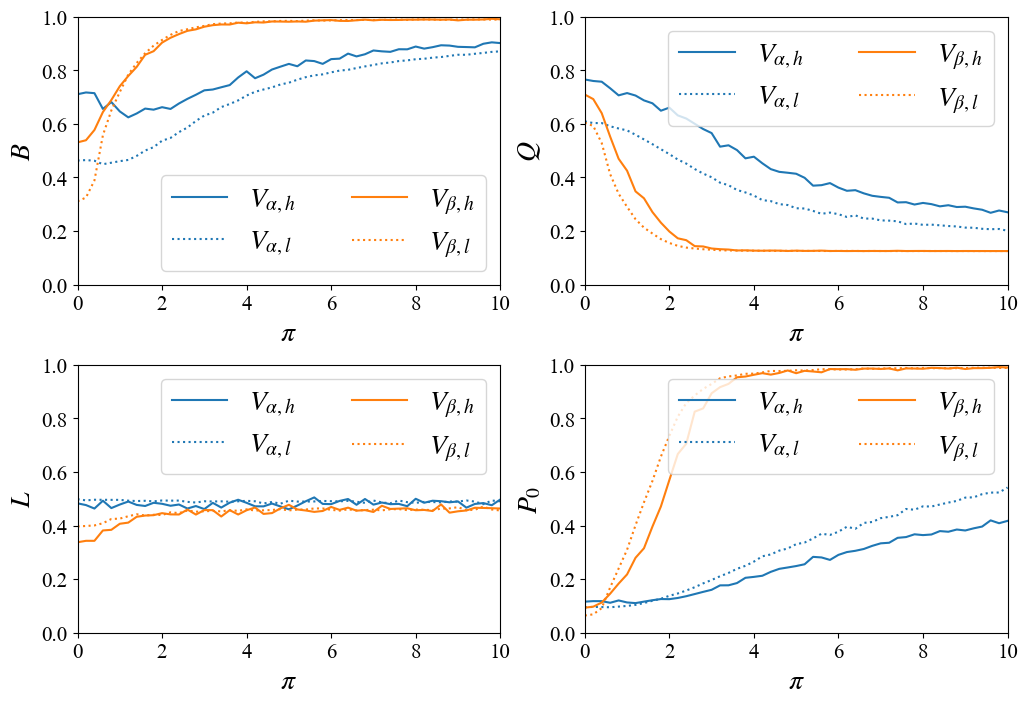}
  \subcaption{S1 (when posting an item)}
  \label{fig:S1param}
\end{minipage}
\\[5pt]
\begin{minipage}[t]{0.95\linewidth}
  \centering
  \includegraphics[width=0.95\linewidth]{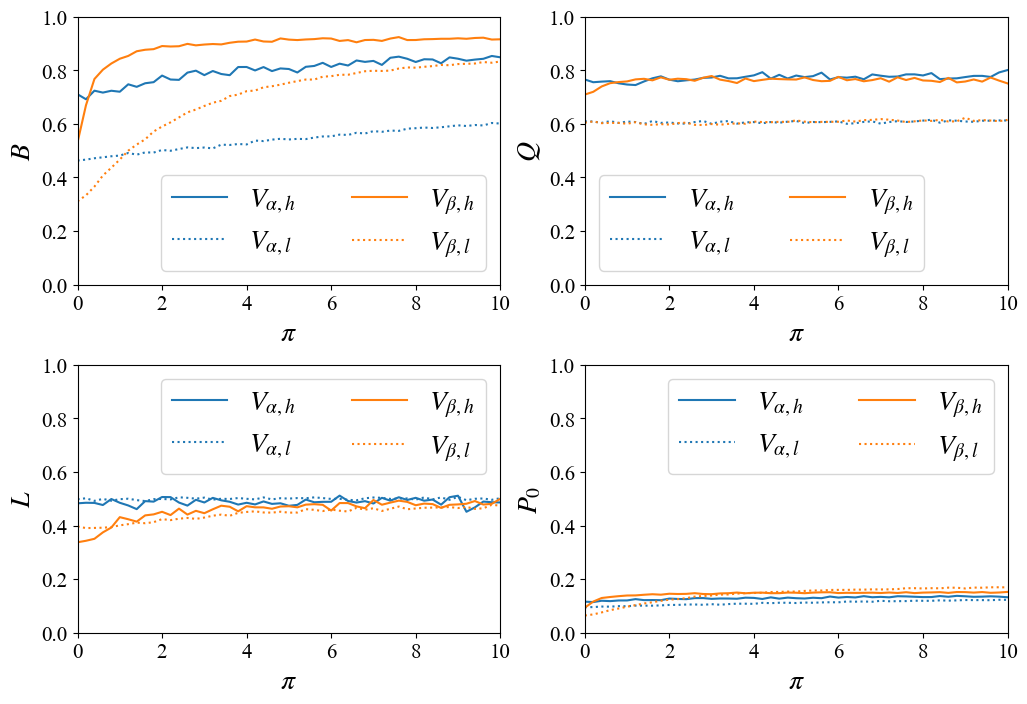}
  \subcaption{S2 (when viewing an item)}
  \label{fig:S2param}
\end{minipage}
\\[5pt]
\begin{minipage}[t]{0.95\linewidth}
  \centering
  \includegraphics[width=0.95\linewidth]{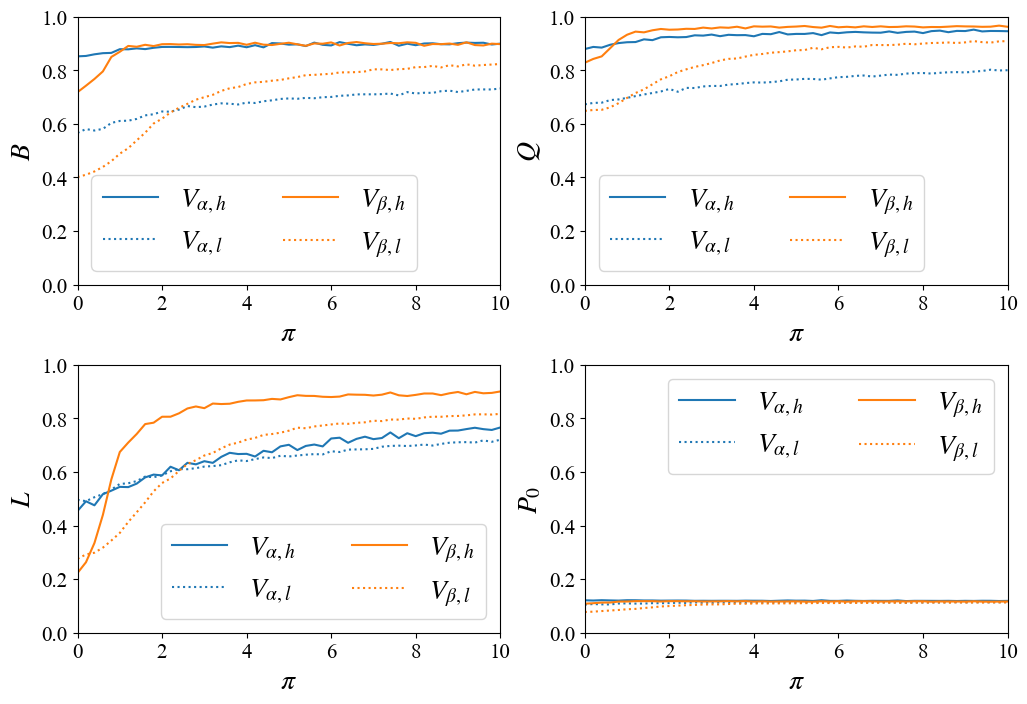}
  \subcaption{S3 (when posting a meta-comment)}
  \label{fig:S3param}
\end{minipage}
\caption{Behavioral strategies of agents}
\label{fig:param}
\end{figure}

\subsection{Strategies under more monetary reward $\pi$}
In the next experiment, we investigated the effect of the monetary
reward value $\pi$ on agents' strategies. For this purpose, we varied
$\pi$ from $0$ (i.e., \Scheme0) to $10.0$ by $0.2$ in schemes
\Scheme1, \Scheme2 and \Scheme3. Subsequently, we divided $V_\alpha$
and $V_\beta$ into subsets;
\[
\begin{aligned}
  V_{\alpha,h}& =\{a\in V_\alpha \mid \deg(a) \geq 50\},
  V_{\alpha,l}=V_{\alpha}\setminus V_{\alpha,h}\\
  V_{\beta,h}& =\{a\in V_\beta \mid \deg(a) \geq 50\}, \textrm{ and }
  V_{\beta,l}=V_{\beta}\setminus V_{\beta,h},
\end{aligned}
\]
where $\deg(a)$ is the degree of agent $a\in A$. The results of the
experiment are plotted in Fig.~\ref{fig:param}, where each plot is the
average during the last (so $1000$th) generation of MWGA. Note that
the rate $P_0$ (Eq.~\ref{eq:post}) is the actual rate of posting after
considering the effort of polishing for quality $Q$.
\par

In scheme \Scheme1, the larger $\pi$ resulted in excess low-quality
items in CGM. Figure \ref{fig:S1param} showed that $B$ and $Q$ of
$V_\beta$ were almost $1.0$ and $0.125$ ($=Q_{min}$) when $\pi \geq
4.0$, and $P_0=1.0$ in this situation under \Scheme1. $B$ and $Q$ of
agents in $V_\alpha$ did not assume such extreme values, but they also
have excess low-quality items, although agents with higher degrees
slightly resisted this tendency. Meanwhile, the value of $L$ was not
significantly affected.
\par

In \Scheme2, all types of agents, particularly those in $V_{\beta,h}$,
increased $B$ with the increase of $\pi$, but variations of $L$, $Q$
and $P_0$ were small as shown in Fig.~\ref{fig:S2param}. Only the
value of $B$ increased because agents increased the chances of being
viewed by neighboring agents while reducing the number of comments
that did not involve monetary rewards. This scheme did not reduce the
quality of the posted items, but neither did it improve them. In
contrast, in scheme \Scheme3, all types of agents increased all $B$,
$L$, and $Q$, although the actual posting rate $P_0$ remained
constant. To gain monetary rewards, the agents that posted items post
meta-comments after receiving the associated comments. Thus, all
agents had to maintain higher comment rates and quality. Furthermore,
the incentive for posting meta-comments relies only on monetary
rewards as meta-comments did not provide psychological rewards.
\par

\begin{figure}[bt]
\centering
\includegraphics[width=0.95\linewidth]{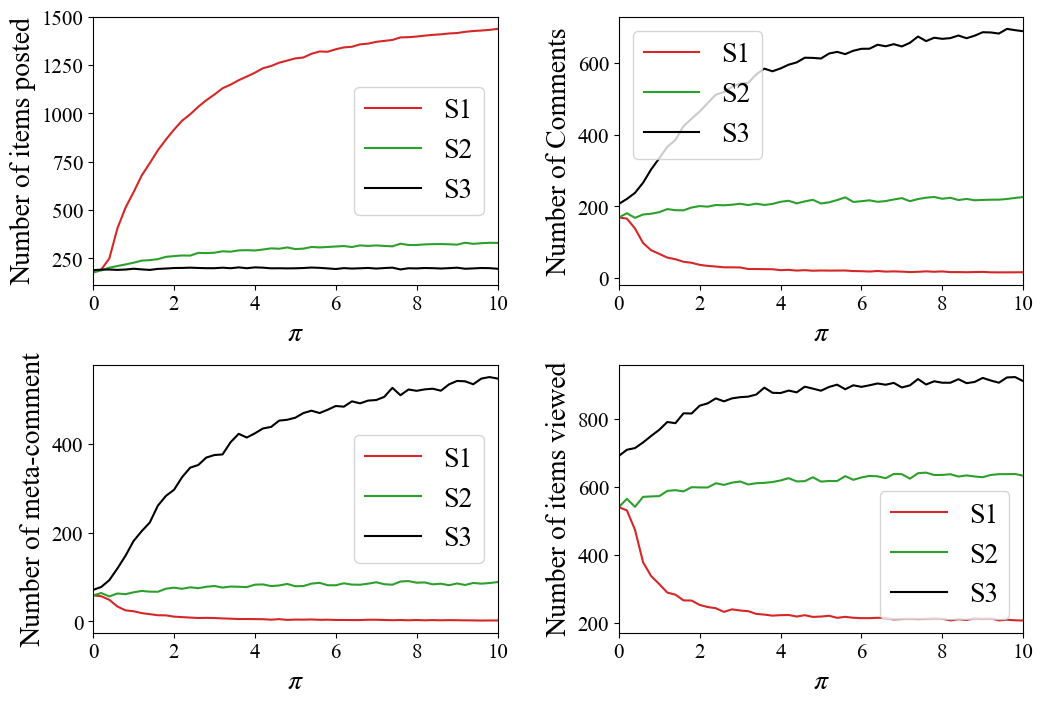}
\caption{Number of objects on CGM}
\label{fig:numOfs}
\end{figure}

\subsection{Analysis of agent activities and $\pi$-effectiveness}
Finally, we counted the number of posted items, comments, and
meta-comments under all monetary reward schemes during the last
generation. The results are plotted in Fig.~\ref{fig:numOfs}. It shows
different trends depending on the schemes. In \Scheme1, only the
number of items increased, but as they were low-quality, other
activities decreased; this seems an undesirable situation. In
\Scheme2, agents increased only slightly in the numbers of all
activities. In contrast, \Scheme3 has a significant effect on the
numbers of viewed items, comments, and meta-comments for an increase
in $\pi$, although the number of posted items with high quality
remained nearly constant.
\par

In addition, as Figs.~\ref{fig:param} and \ref{fig:numOfs} did not
indicate the number of monetary rewards offered, we calculated the
{\em $\pi$-effectiveness} $E_{\it act}^{\Scheme n}(\pi)$, that is, the
relationship between the impact on activities and all rewards
offered. It is defined as
\begin{equation}
E^{\Scheme n}_{\it act}(\pi) = \frac{\mathcal{N}^{\Scheme n}_{\it
    act}(\pi)-\mathcal{N}_{\it act}^{\Scheme n}(0)}{\overline{K}^{\Scheme n}(\pi)},
\end{equation}
where $\Scheme n$ denotes one of schemes ($n=1,2$ or $3$),
$\overline{K}^{\Scheme n}(\pi)$ is the average monetary rewards
offered per agent and $\mathcal{N}^{\Scheme n}_{\it act}(\pi)$ is the
number of activity, {\it act}, under the specified scheme and the
value of $\pi$ during the $1000$-th generation. Therefore, {\it act}
is one of {\it item}, {\it view}, {\it comm}, and {\it meta}, which
correspond to the activities, item post, item view, comment post, and
meta-comment post, respectively. Thus, $\pi$-effectiveness indicates
the improvement by the monetary reward scheme from when \Scheme0 was
adopted.
\par

\begin{figure}[bt]
\centering
\includegraphics[width=0.98\linewidth]{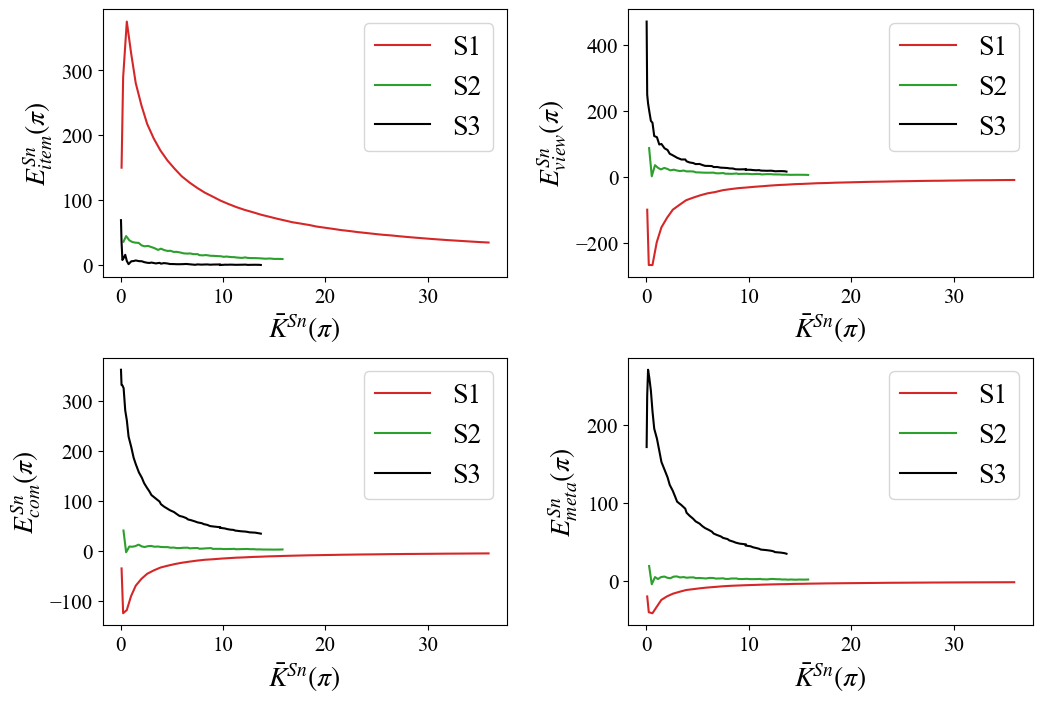}
\caption{Effects of objects on each scheme}
\label{fig:numOfsE}
\end{figure}

The relationship between $\overline{K}^{\Scheme1}(\pi)$ and $E_{\it
  act}^{\Scheme n}(\pi)$ are plotted in Fig.~\ref{fig:numOfsE}. First,
we can see that $\overline{K}^{\Scheme1}(\pi)$ became larger (over
$30$) than others. Because $\overline{K}^{\Scheme1}(\pi)\times N$ is
the total monetary rewards offered to all agents, it means that
numerous monetary rewards are offered to users. It also shows that
$E^{\Scheme1}_{\it item}(\pi)$ sharply increased and then quickly
decreased according to the increase of $\overline{K}^{\Scheme
  n}(\pi)$. In \Scheme3, $E^{\Scheme3}_{\it view}(\pi)$,
$E^{\Scheme3}_{\it comm}(\pi)$ $E^{\Scheme3}_{\it meta}(\pi)$ raised
and then also decreased. From these results and Fig.~\ref{fig:param},
we can say that $\pi$ did not need to be high. The larger $\pi$
quickly reduced the effectiveness in terms of improving all types of
activities and the quality of items. 
\par

\section{Conclusion}
In this study, we investigated the impact of monetary rewards on
agents' activities and how differently agents' activities are affected
depending on their places in a CGM network. For this purpose, we
improved the SNS-NG/MQ by eliminating unnecessary parts and then
identified the agents' appropriate behaviors using MWGA. We adopted
three monetary reward schemes and discovered that they affect agents'
behaviors differently. Our results indicate that the agents with high
degrees were more likely to be affected by certain monetary reward
schemes and by appropriately setting the schemes, all agents can
improve the quality of items they post without reducing the frequency
of posting despite the cost of quality items. Finally, while higher
rewards led to increased activity, they were not typically necessary
from the perspective of the effectiveness of monetary rewards.
Our model is useful for analyzing rational strategies for individual 
standpoints for specific schemes of monetary rewards by setting 
appropriate reward values, and it is also useful for identifying optimal
 schemes in various media.
\par

Thus far, we have only addressed unlimited monetary rewards for users
provided a media platformer, but this is unrealistic although our
results suggested that no high rewards were needed. Future work can
focus on an analysis using a model that either places realistic limits
on the media's monetary reward scheme based on the budgets or
considers incentives for advertising and revenue generation such as
memberships.

\bibliographystyle{unsrt}
\bibliography{ref.bib}

\end{document}